\documentclass[a4paper,preprintnumbers,amsmath,amssymb,nofootinbib,10pt]{revtex4}

\usepackage[margin=2cm]{geometry}
\usepackage{amssymb}

\def\beq{\begin{equation}}
\def\eeq{\end{equation}}
\def\baq{\begin{eqnarray}}
\def\eaq{\end{eqnarray}}

\def\bea{\begin{eqnarray}}
\def\eea{\end{eqnarray}}
\def\be{\begin{equation}}
\def\ee{\end{equation}}

\def\Tvac{\check{T}}
\def\rhovac{\check{\rho}}
\def\Pvac{\check{P}}
\def\uvac{\check{u}}
\def\vvac{\check{v}}
\def\thetavac{\check{\theta}}
\def\zetavac{\check{\zeta}}
\def\Svac{\check{S}}

\def\de{{\rm de}}
\def\gcg{{\rm gCg}}

\begin{document}

\title{Inhomogeneous vacuum energy}


\author{{David Wands$^{\,a}$, Josue De-Santiago$^{\,a,b}$ and Yuting Wang$^{\,a,c}$}\\
%
$^a$Institute of Cosmology $\&$ Gravitation, University of Portsmouth, Dennis Sciama Building,\\\hskip0.2cm
Portsmouth, PO1 3FX, United Kingdom
\\ \hskip0.2cm
$^b$Universidad Nacional Aut\'onoma de M\'exico, 04510, D.~F., M\'exico
\\ \hskip0.2cm
$^c$School of Physics and Optoelectronic Technology, Dalian University of Technology, Dalian,\\\hskip0.2cm
Liaoning 116024, People's Republic of China
\\
}
\date{April 16, 2012}

\begin{abstract}
Vacuum energy remains the simplest model of dark energy which could drive the accelerated expansion of the Universe without necessarily introducing any new degrees of freedom. Inhomogeneous vacuum energy is necessarily interacting in general relativity. Although the four-velocity of vacuum energy is undefined, an interacting vacuum has an energy transfer and the vacuum energy defines a particular foliation of spacetime with spatially homogeneous vacuum energy in cosmological solutions. It is possible to give a consistent description of vacuum dynamics and in particular the relativistic equations of motion for inhomogeneous perturbations given a covariant prescription for the vacuum energy, or equivalently the energy transfer four-vector, and we construct gauge-invariant vacuum perturbations. We show that any dark energy cosmology can be decomposed into an interacting vacuum+matter cosmology whose inhomogeneous perturbations obey simple first-order equations.
\end{abstract}

\maketitle


\section{Introduction}

The apparent acceleration of the Universe today has led cosmologists to consider a great variety of different models for the dark energy which apparently dominates the expansion. Vacuum energy is the simplest model of dark energy. If spacetime retains a non-zero energy density even in the absence of any particles then this energy density would be undiluted by the cosmological expansion and could drive an accelerated expansion as the density of ordinary matter or radiation becomes sub-dominant. The simplest example of constant vacuum energy density, $V$, in Einstein gravity is equivalent to a cosmological constant, $\Lambda=8\pi G_N V$, and the discrepancy between the value of the energy density required by current observations \cite{Copeland:2006wr} and the typical energy scales predicted by particle physics is the long-standing cosmological constant problem \cite{Weinberg:1988cp}.

The idea of decaying vacuum energy is a recurring concept in attempts to explain the present acceleration \cite{Bertolami:1986bg,Freese:1986dd,Chen:1990jw,Carvalho:1991ut,Berman:1991zz,Pavon:1991uc,AlRawaf:1995rs,Shapiro:2000dz,Sola:2011qr}. Unlike other models of dark energy, vacuum energy does not introduce any new dynamical degrees of freedom. It is the energy density of nothing.
There seems to be little to be learnt from simply assuming an arbitrary time-dependent vacuum to obtain the desired cosmological solution. Ideally one should have a physical model from which one could not only derive the time-dependent solution but also study other physical effects, including the evolution of inhomogeneous perturbations which can be tested against other cosmological observations.
For example, vacuum fluctuations of free fields can support an averaged density proportional to the fourth-power of the Hubble expansion, $V\propto H^4$ \cite{Bunch:1978yq}. Such a vacuum energy would not in itself support an accelerated expansion, but other forms such as $V\propto H$ have been proposed \cite{Schutzhold:2002pr} better able to match the observational data \cite{Fabris:2006gt,VelasquezToribio:2009qp,Zimdahl:2011ae,Xu:2011qv,Alcaniz:2012mh}. It is not clear whether or how such time-dependent vacuum models can be compared with observations in an inhomogeneous universe. In this paper we will argue that it is possible to give a consistent description of vacuum dynamics, and in particular the relativistic equations of motion for inhomogeneous perturbations, given a covariant, physical prescription for the local vacuum energy or, equivalently, the vacuum energy transfer 4-vector, $Q_\mu=-\nabla_\mu V$. Thus one should be able to subject vacuum models to observational constraints even in the absence of a Lagrangian derivation or microphysical description.

One class of dark energy models that do have a well-defined microphysical prescription are quintessence models where the vacuum energy is determined by the self-interaction potential energy of one (or more) scalar fields, $V(\varphi)$ \cite{Ratra:1987rm,Caldwell:1997ii,ArmendarizPicon:2000ah}. A canonical scalar field has both potential and kinetic energy, therefore the vacuum potential energy interacts with the kinetic energy of the scalar field \cite{Malik:2004tf}. The introduction of a scalar field leads to a new degree of freedom and one can recover the Klein-Gordon equations of motion for the field from energy conservation, allowing for the energy transfer between the scalar field and the vacuum.
Quintessence has been generalised to consider scalar fields also interacting with matter, known as coupled quintessence~\cite{Wetterich:1994bg,Amendola:1999er,Holden:1999hm,Billyard:2000bh,Farrar:2003uw,Boehmer:2008av}.

In this paper we will consider a general form for the vacuum energy which may directly interact with existing fields or fluids without invoking any other additional degrees of freedom. We show that any homogeneous dark energy cosmology can be decomposed into an interacting vacuum+matter cosmology. By lifting this spatially-homogeneous solution to a covariant interaction one can study inhomogeneous perturbations which obey coupled first-order equations of motions for the matter density and velocity. We present the equations of motion for linear perturbations in Einstein gravity and identify gauge-invariant variables for the vacuum perturbations. As an example we consider perturbations of a generalised Chaplygin gas cosmology, decomposed into an interacting vacuum+matter.

\section{Vacuum energy-momentum}

We define a vacuum energy $V$ to have an energy-momentum tensor proportional to the metric:
\be
 \label{Tvac}
 \Tvac^\mu_\nu = - V g^\mu_\nu  \,.
 \ee
By comparison with the energy-momentum tensor of a perfect fluid
\be
 \label{Tmunu}
 T^\mu_\nu = P g^\mu_\nu +  (\rho+P)u^\mu u_\nu \,,
 \ee
we identify the vacuum energy density and pressure with $\rhovac=-\Pvac=V$. Since $\rhovac+\Pvac=0$ the four-velocity of the vacuum, $\uvac^\mu$, is undefined, as expected in the absence of any particle flow.

Note that the energy density and four-velocity of a fluid can be identified with the eigenvalue and eigenvector of the energy-momentum tensor (\ref{Tmunu}):
\be
 T^\mu_\nu u^\nu= -\rho u^\mu \,.
 \ee
Because the vacuum energy-momentum tensor (\ref{Tvac}) is proportional to the metric tensor, any four-velocity, $u^\mu$, is an eigenvector
\be
 \Tvac^\mu_\nu u^\nu= - V u^\mu \quad \forall \ u^\mu \,,
 \ee
and all observers see the same vacuum energy density, $V$, i.e., the vacuum energy is boost invariant.

However there may be an energy transfer associated with the vacuum energy. Let
\be
\label{vac-conservation}
 \nabla_\mu \Tvac^\mu_\nu = Q_\nu \,.
 \ee
For the vacuum energy-momentum tensor (\ref{Tvac}), since the metric tensor is covariantly conserved, we have
\be
 Q_\nu = - \nabla_\nu V \,.
 \ee
We see that if the vacuum energy is covariantly conserved, $Q_\nu=0$, then the vacuum energy must be homogeneous in spacetime, i.e., $\nabla_\nu V=0$. A non-interacting vacuum energy is equivalent to a cosmological constant in Einstein gravity. Conversely, an interacting vacuum, $Q_\nu\neq0$, is inhomogeneous in spacetime.

Although the vacuum does not have a unique four-velocity, we can use the energy flow, $Q_\nu$, to define a preferred unit four-vector in an interacting vacuum
\be
 \label{def-uvac}
 \uvac^\mu = \frac{-\nabla^\mu V}{|\nabla_\nu V\nabla^\nu V|^{1/2}} \,.
 \ee
normalised such that $\uvac_\mu\uvac^\mu=\pm1$ for a spacelike or timelike flow. Note however that the $\uvac^\mu$ defines a potential flow, i.e, with vanishing vorticity: $\nabla_{[\mu}\uvac_{\nu]}=0$

Assuming that the total energy-momentum for ordinary matter plus the vacuum, $T^\mu_\nu+\Tvac^\mu_\nu$, is covariantly conserved (as is required by the Ricci identity in Einstein gravity) we therefore have
\be
 \label{cov-matter-cons}
 \nabla_\mu T^\mu_\nu = - Q_\nu \,.
 \ee
One might consider a simple case where the energy transfer is proportional to the matter four-velocity, $Q_\mu=Qu_\mu$ or equivalently $\uvac_\mu=u_\mu$, but in fact this is a restrictive assumption since the energy flow (\ref{def-uvac}) is then a potential flow meaning that it cannot too admit any vorticity. In this case the matter velocity itself would be described by a scalar field, $u_\mu\propto \nabla_\mu V$.

\section{FRW cosmology}

\subsection{Spatially homogeneous background}

The Einstein equations for a spatially homogeneous and isotropic Friedmann-Robertson-Walker (FRW) cosmology, with scale factor $a$ and Hubble rate $H=\dot{a}/a$, reduce to the Friedmann equation
\be
 H^2 = \frac{8\pi G_N}{3} \left( \rho + V \right) - \frac{K}{a^2} \,,
 \ee
and the continuity equations for vacuum and matter
\bea
 \dot\rho + 3H(\rho+P) &=& -Q
\,,\\
 \dot{V} &=& Q
\,.
\eea

In this case the vacuum and matter are both spatially homogeneous on spatial hypersurfaces orthogonal to the matter four-velocity, $u^\mu$. The vacuum energy is undiluted by the cosmological expansion, but does have a time-dependent density in the presence of a non-zero energy transfer, $Q\neq 0$, where $Q_\nu=Q u_\nu$. Note that the energy flow, $\uvac^\mu$, and matter velocity, $u^\mu$, necessarily coincide in FRW cosmology due to the assumption of isotropy.

\subsection{Linear perturbations}

For simplicity we will consider linear, scalar perturbations about a spatially flat ($K=0$) FRW metric \cite{Kodama:1985bj,Mukhanov:1990me,Malik:2004tf,Malik:2008im}, with the line element:
\be
 ds^2 = -(1+2\phi)dt^2+2a\partial _i B dt dx^i
  + a^2 \left[(1-2\psi)\delta_{ij}+2\partial_i\partial_j E \right] dx^i dx^j \,.
 \ee
The energy and pressure of matter is given by $\rho+\delta\rho$ and $P+\delta P$, and the four-velocity of matter is given by
\be
 \label{u}
 u^\mu = \left[ 1-\phi \,, a^{-1}\partial^i v \right] \,, \quad u_\mu = \left[ -1-\phi \,, \partial_i \theta \right] \,.
 \ee
where we define $\partial^iv=a(\partial x^i/\partial t)$ and $\theta=a(v+B)$.

Once we allow for deviations from homogeneity in the matter and metric, we must also allow for inhomogeneity in an interacting vacuum. As remarked earlier, the vacuum has an energy density and pressure, but no unique velocity. In particular the momentum of vacuum vanishes in any frame, $(\rhovac+\Pvac)\theta=0$. On the other hand the energy flow $\uvac$ defined in Eq.~(\ref{def-uvac}), can be written in analogy with the fluid velocity (\ref{u}) as
\be
 \uvac^\mu = \left[ 1-\phi \,, a^{-1}\partial^i \vvac \right] \,, \quad \uvac_\mu = \left[ -1-\phi \,, \partial_i \thetavac \right] \,.
 \ee
where from Eq.~(\ref{def-uvac}) we identify $\thetavac=-\delta V/\dot{V}$.

The energy continuity equations for matter and vacuum become
\bea
 \dot{\delta\rho} + 3H(\delta\rho+\delta P)
 -3 (\rho+P) \dot\psi + (\rho+P)\frac{\nabla^2}{a^2} \left( \theta + a^2\dot{E} - aB \right)
 &=& -\delta Q - Q\phi
 \,,\\
 \dot{\delta V} &=& \delta Q + Q\phi \,.
\eea
while the momentum conservation becomes
\bea
(\rho+P)\dot\theta - 3c_s^2H(\rho+P)\theta + (\rho+P)\phi
+\delta P \nonumber &=& - f + c_s^2 Q \theta
 \,,\\
 -\delta V &=& f + Q\theta \,.
\eea
where the adiabatic sound speed $c_s^2\equiv \dot{P}/\dot\rho$ and following \cite{Kodama:1985bj,Malik:2004tf,Malik:2008im} we decompose
\be
 Q_\mu = \left[ -Q(1+\phi)-\delta Q \,, \partial_i (f + Q\theta) \right] \,.
 \ee
Note that the vacuum momentum conservation equation becomes a constraint equation which requires that the vacuum pressure gradient, $\partial_i(-\delta V)$ is balanced by the force $\partial_i(f+Q\theta)$. This determines the equal and opposite force exerted by the vacuum on the matter:
\be
- f = \delta V + \dot{V}\theta \,.
 \ee
i.e., the fluid element feels the gradient of the vacuum potential energy.
We see explicitly that perturbations of a fluid coupled to the vacuum with $\Pvac=-\rhovac$ has no additional degrees of freedom, in contrast to a dark energy fluid with $P_X\neq -\rho_X$. Using the vacuum energy and momentum conservation equations to eliminate $\delta Q$ and $f$ we obtain
\bea
 \label{finaldeltarho}
 \dot{\delta\rho} + 3H(\delta\rho+\delta P)
 -3 (\rho+P)\dot\psi + (\rho+P) \frac{\nabla^2}{a^2} \left( \theta + a^2\dot{E} - aB \right)
 &=& - \dot{\delta V}
 \,,\\
 \label{finaltheta}
(\rho+P)\dot\theta - 3c_s^2H(\rho+P)\theta + (\rho+P)\phi
+\delta P &=& \delta V + (1+c_s^2) \dot{V} \theta
 \,.
\eea

The metric perturbations obey the Einstein constraint equations
\bea
 \label{energy-con}
3H\left( \dot\psi +H\phi \right) - \frac{\nabla^2}{a^2} \left[ \psi + H \left( a^2\dot{E} - aB \right) \right] = -4\pi G \left( \delta\rho + \delta V \right)
 \,,
\\
 \label{mtm-con}
\dot\psi +H\phi = - 4\pi G (\rho+P) \theta
 \,.
 \eea
As expected, the vacuum contributes to the energy constraint (\ref{energy-con}) but not the momentum constraint (\ref{mtm-con}).

In summary, we have seven first-order scalar perturbations, $\phi$, $\psi$, $E$, $B$, $\delta\rho$, $\theta$ and $\delta V$. We can eliminate two variables by choice of temporal and spatial gauge, and we have the two Einstein constraint equations (\ref{energy-con}) and (\ref{mtm-con}). This leaves three variables to be determined from the two first-order energy and momentum conservation equations (\ref{finaldeltarho}) and (\ref{finaltheta}). Thus we also need a physical model to describe the vacuum interactions, $\delta V$, in order to obtain a closed system of equations.

\subsection{Gauge invariant perturbations}

It is well known that metric and matter perturbations can be gauge-dependent. Under a first-order gauge transformation, such as $t\to t+\delta t(t,x^i)$, the fluid density and pressure transform as $\delta\rho\to\delta\rho-\dot\rho\delta t$ and $\delta P\to \delta P-\dot{P}\delta t$. However a quantity such as the non-adiabatic pressure perturbation, $\delta P_{\rm nad}=\delta P -(\dot{P}/\dot\rho)\delta\rho$, is gauge-invariant.

Similarly the vacuum perturbation transforms as $\delta V\to \delta V -Q\delta t$ and $\delta Q\to \delta Q-\dot{Q}\delta t$.
Given $\delta\Pvac=-\delta\rhovac=-\delta V$ and $\dot{\Pvac}=-\dot{\rhovac}=-Q$ it is clear that the intrinsic non-adiabatic pressure perturbation of the vacuum is zero, as it is for any barotropic fluid with $P=P(\rho)$.

The fluid 3-momentum transforms as $\theta\to \theta+\delta t$ and the energy flow transforms similarly as $\thetavac\to\thetavac+\delta t$ so that the relative velcity $\thetavac-\theta$ is gauge invariant.
The comoving-orthogonal density perturbation (on hypersurfaces orthogonal to the fluid four-velocity $u^\mu$) is thus given by the gauge-invariant combination
\be
\delta\rho_{\rm com} = \delta\rho + \dot\rho\theta \,.
\ee
Therefore the comoving orthogonal vacuum density perturbation is
\be
 \label{defdeltarhovac}
 \delta\rhovac_{\rm com} = \delta V + \dot{V} \theta \,.
 \ee
This is in general non-zero, but the vacuum perturbation on hypersurfaces orthogonal to the energy flow $\uvac^\mu$ is identically zero:
\be
 \Delta\rhovac_{\rm com} = \delta V + \dot{V} \thetavac = 0 \,,
 \ee
given $\thetavac=-\delta V/\dot{V}$. In fact this is a consequence of the energy flow being the gradient of the vacuum energy, as given in Eq.~(\ref{def-uvac}), and therefore the orthogonal hypersurfaces are uniform vacuum energy hypersurfaces by construction. Note that we can therefore write the comoving vacuum density perturbation (\ref{defdeltarhovac}) as
\be
 \label{defdeltarhovac2}
 \delta\rhovac_{\rm com} = \dot{V} \left( \theta - \thetavac \right) \,.
 \ee
Therefore, if the energy flow follows the fluid four-velocity, $\uvac^\mu=u^\mu$, then we have $\thetavac=\theta$ and the vacuum is spatially homogeneous on comoving-orthogonal hypersurfaces, $\delta\rhovac_{\rm com}=0$.

The gauge-invariant vacuum density perturbation on spatially-flat hypersurfaces is given by
\be
 \label{defzetavac}
 \zetavac = -\psi - \frac{H}{\dot{V}} \delta V \,,
 \ee
and this will in general be non-zero. Another gauge invariant expression for the vacuum density perturbation is the vacuum density perturbation on uniform-fluid density hypersurfaces, which describes a relative density perturbation
\be
 \label{defSvac}
 \Svac = 3 \left(\zetavac - \zeta \right)
 = -3H \left( \frac{\delta V}{\dot{V}} - \frac{\delta\rho}{\dot\rho} \right) \,.
 \ee
Therefore, if the vacuum energy is a function of the local matter density, $V=V(\rho)$, then the relative density perturbation must vanish and the vacuum is spatially homoegeneous on uniform-density hypersurfaces, $\Svac=0$.

The total non-adiabatic pressure perturbation due to any intrinsic non-adiabatic pressure of the fluid and the relative entropy perturbation between the vacuum and the fluid is then
\be
 \label{defPnad}
 \delta P_{\rm nad}
  = \delta P - c_s^2 \delta\rho + \frac{(1+c_s^2)Q[Q+3H(\rho+P)]}{9H^2(\rho+P)} \Svac \,.
 \ee
This vanishes for adiabatic fluid perturbations, $\delta P=c_s^2\delta\rho$, and adiabatic vacuum fluctuations, $\Svac=0$, or a non-interacting vacuum, $Q=0$.

\section{Dark energy models}

\subsection{Scalar field dynamics}

As remarked earlier, a self-interacting scalar field, $\varphi$, has a vacuum energy, $V(\varphi)$, which may be inhomogeneous if it interacts with the kinetic energy of the scalar field.
Consider a single scalar field with Lagrangian density ${\cal L}=F(X)-V(\varphi)$ where $X=-g^{\mu\nu}\nabla_\mu\varphi\nabla_\nu\varphi/2$ \cite{ArmendarizPicon:1999rj,ArmendarizPicon:2000ah,DeSantiago:2012inprep}.
In an FRW cosmology, a homogeneous scalar field with vanishing potential ($V=0$) has kinetic energy density $\rho_F=2XF_{,X}-F$ and pressure $P_F=F$ \cite{Scherrer:2004au}. In particular for $F(X)\propto X^n$ we have a barotropic fluid with constant equation of state $w_F=P_F/\rho_F=(2n-1)^{-1}$ \cite{Garriga:1999vw}, which equals one for a canonical field, $F=X$. The energy flow from kinetic to potential energy, Eq.(\ref{vac-conservation}), is given by $Q_\mu=-(\partial V/\partial\varphi)\nabla_\mu\varphi$ \cite{Malik:2004tf}.

The energy continuity equation (\ref{cov-matter-cons}) for the scalar field kinetic energy requires
\be
 \nabla_\mu T^\mu_\nu = \left( F_{,X}\Box\varphi + F_{,XX}\nabla_\mu\varphi\nabla^\mu X \right) \nabla_\nu\varphi = -Q_\nu
 \,.
 \ee
Hence we recover the generalised Klein-Gordon equation \cite{ArmendarizPicon:2000ah}
\be
 F_{,X}\Box\varphi + F_{,XX}\nabla_\mu\varphi\nabla^\mu X - \frac{\partial V}{\partial\varphi} = 0 \,.
 \ee

\subsection{Decomposing dark energy models}

Numerous dark energy cosmologies have been proposed in terms of exotic dark energy solutions $\rho_\de(a)$ or equations of state $P_\de(\rho_\de)$ and/or adiabatic sound speed $c_s^2$. Some of these models have been proposed as unified dark energy models \cite{Kamenshchik:2001cp} which can play the role of both dark matter ($P_\de\approx0$) and vacuum energy ($P_\de\approx-\rho_\de$) at different stages of cosmological evolution.

Any
dark energy fluid energy-momentum tensor (\ref{Tmunu}) with density $\rho_\de$ can be described by pressureless matter, with density $\rho_m$ and velocity $u^\mu_m=u^\mu$, interacting with the vacuum, $V$, such that $\rho_\de=\rho_m+V$. The corresponding matter and vacuum densities are given by
\be
 \label{decompose}
 \rho_m = \rho_\de+P_\de \,, \quad V = -P_\de \,.
 \ee
while the energy flow is $Q_\mu=\nabla_\mu P_\de$.
In an FRW cosmology this corresponds to $Q=-\dot{P}_\de$.
One might choose to decompose a dark energy model $\rho_\de(a)$ into any two interacting barotropic fluids such that $\rho_\de=\rho_1+\rho_2$, but this would double the degrees of freedom in the model unless one of these two ``fluids'' is the vacuum.

Linder and Scherrer \cite{Linder:2008ya} showed that any dark energy cosmology could be represented by a constant vacuum energy, $V$, and a fluid with an suitable equation of state, $w_f=(P_{de}+V)/(\rho_{de}-V)$. In this case the vacuum energy was non-interacting but the fluid required a specific equation of state, whereas in our case we will consider pressureless matter, but the vacuum then requires a specific interaction.\footnote{In practice one could choose any equation of state for the matter and a specific interaction such that $\rho_m+V=\rho_{de}$. An example is a quintessence model, where a canonical scalar field has sound speed equal to one, which reproduces any given dark energy cosmology for a suitably chosen potential.} These are both examples of so-called dark degeneracy~\cite{Kunz:2007rk,Aviles:2011ak}.

One much studied dark energy model is the generalised Chaplygin gas, defined by the barotropic equation of state \cite{Kamenshchik:2001cp,Bento:2002ps}
\be
 \label{Pgcg}
 P_\gcg = -A \rho_\gcg^{-\alpha} \,.
 \ee
This leads to a cosmological solution for the density
\be
 \rho_\gcg = \left( A + Ba^{-3(1+\alpha)} \right)^{1/(1+\alpha)} \,.
 \ee
This has the simple limiting behaviour $\rho_\gcg\propto a^{-3}$ as $a\to0$ and $\rho_\gcg\to A^{1/(1+\alpha)}$ as $a\to+\infty$, therefore this has been proposed as a unified dark matter model. However such models are strongly constrained by observations since the barotropic equation of state defines a sound speed for matter perturbations which only reproduces the successful $\Lambda$CDM model when $\alpha\to0$ \cite{Sandvik:2002jz,Park:2009np}.

The decomposition (\ref{decompose}) into pressureless matter interacting with the vacuum has previously been considered for the generalised Chaplygin gas by Bento et al \cite{Bento:2004uh}. In this case
we have the FRW solution
 \be
 V =
 A \left( A + Ba^{-3(1+\alpha)} \right)^{-\alpha/(1+\alpha)}
   \,,
\ee
and hence
\be
\label{defA}
 A = (\rho_m + V)^\alpha V \,.
 \ee
The form of the FRW solution suggests a simple interaction
\be
 \label{Qgcg}
 Q
= 3 \alpha H \left( \frac{\rho_m V}{\rho_m + V} \right) \,.
 \ee
In the matter or vacuum dominated limits this reduces to an interaction of the form $Q\propto H\rho_m$ or $Q\propto HV$ studied, for example, by Barrow and Clifton~\cite{Barrow:2006hia}.

It is intriguing to note that the FRW model can be defined in terms of an interaction (\ref{Qgcg}) with a single dimensionless parameter, $\alpha$, whereas when defined in terms of an equation of state (\ref{Pgcg}) its definition requires both $\alpha$ and the dimensional constant $A$ which determines the late-time cosmological constant. In the interaction model, $A$, and therefore the late-time cosmological constant, emerges as an integration constant dependent on initial conditions.

However, to study perturbations in the decomposed model we must ``lift'' the explicitly time-dependent FRW solution to a covariant model for the interaction. We appear to have (at least) two choices.

Firstly we could require that the energy flow follows the matter four-velocity, $Q_\mu=Qu_\mu$, such that $\thetavac=\theta$.
%
In any interacting matter+vacuum model where the energy flow, $Q_\mu$, is along the matter four-velocity, $u_\mu$, the perturbation equations (\ref{finaldeltarho}) and (\ref{finaltheta}) in an arbitrary gauge reduce to the equations for the matter density and velocity perturbation
\bea
 \label{deltarhom}
 \dot{\delta\rho}_m + 3H\delta\rho_m
 -3 \rho_m \dot\psi + \rho_m \frac{\nabla^2}{a^2} \left( \theta + a^2\dot{E} - aB \right)
 &=& - \dot{\delta V}
 \,,\\
  \label{thetam}
\dot\theta + \phi &=& 0
 \,.
\eea
Because $\thetavac=\theta$, the comoving density perturbation for the vacuum (\ref{defdeltarhovac2}) is zero and hence there is no force exerted by the vacuum on the matter particles in Eq.(\ref{finaltheta}).
There is no pressure perturbation in the comoving gauge and hence the sound speed for the interacting matter+vacuum is zero, $c^2=0$.

These equations are simplest in a comoving synchronous gauge where $\theta=0$ and $\phi=0$. Eq.~(\ref{thetam}) is automatically satisfied and the Einstein momentum constraint (\ref{mtm-con}) then requires that $\dot\psi=0$, i.e., $\psi=C(x^i)$. Since we have $\thetavac=0$ in the comoving gauge, we also have $\delta V=0$. Thus the matter density perturbation equation (\ref{deltarhom}) and Einstein energy constraint (\ref{energy-con}) are the same as in the absence of any vacuum energy
\bea
 \dot{\delta\rho}_m + 3H\delta\rho_m
 + \rho_m \frac{\nabla^2}{a^2} \left( a^2\dot{E} - aB \right)
 &=& 0
 \,,\\
  \frac{\nabla^2}{a^2} \left[ \psi + H \left( a^2\dot{E} - aB \right) \right] &=& 4\pi G \delta\rho_m
 \,,
\eea
except that the background evolution is that for an interacting vacuum+matter cosmology.

These perturbation equations are strikingly simple, apparently the same as if we had treated the vacuum as spatially homogeneous from the start, but in fact these equations do allow a gauge-invariant vacuum perturbation, $\zetavac=-C(x^i)$, and hence a relative entropy perturbation, $\Svac$.
{}From Eqs.~(\ref{defSvac}) and (\ref{defA}) we can identify
\be
 \Svac = - \frac{(\rho_m+V)^2}{\alpha\rho_m\left(\rho_m+(1+\alpha)V\right)}\frac{\delta A}{A} \,,
 \ee
Hence from Eq.~(\ref{Pgcg}) we can identify the relative entropy perturbation with a change in the local equation of state and non-adiabatic pressure perturbation (\ref{defPnad}), $\delta P_{\rm nad}=-V\delta A/A$.

On the other hand the requirement that $Q_\mu=Qu_\mu$ does impose a restriction on the type of matter field that can be coupled to the vacuum. It must be irrotational with $u_\mu\propto \nabla_\mu V$. And we do not have a clear covariant model for the interaction $Q=-u^\mu Q_\mu$. By requiring $\delta V = -Q\theta$, we construct the vacuum perturbation, $\delta V$, iteratively given the matter velocity, $\theta$, and the background energy transfer, $Q$ in Eq.~(\ref{Qgcg}).

Alternatively one might instead define a covariant model for the energy flow, $Q_\mu$, independent of the matter velocity. In this case we need to identify the homogeneous interaction (\ref{Qgcg}) as a total derivative, $Q=\dot{V}$. Given the homogeneous solution (\ref{defA}) the simplest covariant choice appears to be
\be
 \label{Qmugcg}
 Q_\mu = - \nabla_\mu \left( \frac{A}{(\rho_m+V)^\alpha} \right) \,.
\ee
The comoving vacuum perturbation, $\delta\rhovac_{\rm com}$ in Eq.~(\ref{defdeltarhovac2}), is in general non-zero in this model since $\thetavac\neq\theta$.

In an FRW cosmology the interaction (\ref{Qmugcg}) reduces to Eq.~(\ref{Qgcg}) as required if $\dot{A}=0$. However the dimensional constant $A$ now appears again in the covariant defintion of the model, defining the interaction (\ref{Qmugcg}), and hence the late-time cosmological constant. If $A$ is a universal constant, $\nabla_\mu A=0$, then we have $\nabla_\mu V\propto \nabla_\mu\rho_m$ and the relative entropy perturbation (\ref{defSvac}) must vanish, $\Svac=0$. In this case the interacting matter+vacuum, including inhomogeneous perturbations, is exactly equivalent to the barotropic fluid model for the Chaplygin gas (\ref{Pgcg}) with adiabatic sound speed, $c^2=\dot{P}/\dot\rho$.


\section{Conclusions}
\label{sec:conclusions}

In this paper we have considered vacuum energy as a source of spacetime curvature in Einstein gravity. We have shown that an inhomogeneous vacuum energy implies an interacting vacuum. The vacuum energy is a scalar potential and thus defines a particular spacetime foliation which, in a cosmological solution, will be spatially homogeneous.

We have shown that in order to consistently consider inhomogeneous perturbations in the presence of a vacuum energy one must specify a covariant, physical model for the vacuum energy. Simply specifying a particular time-dependence for the vacuum energy in an FRW cosmology is unsatisfactory if the vacuum energy is introduced solely to produce a particular time-dependence of the cosmological expansion. It is more satisfactory to specify a physical model for the vacuum energy, or equivalently its interactions, from which the time-dependent solutions can then be deduced. Even if the time-dependent solutions are no different, one can go on to consider the behaviour of inhomogeneous perturbations which can be compared against further observational constraints.

The familiar example of an interacting vacuum is the potential energy of a scalar field which is inhomogeneous if the potential energy is transferred to the kinetic energy of the field. The equations presented here simply offer a different perspective on the dynamics of a scalar field with vacuum potential energy but do not change the dynamics. On the other hand our equations enable us to consider vacuum energy interacting with other forms of matter, including pressureless matter or radiation. We can therefore consider vacuum energy models which do not introduce any degrees of freedom beyond those already present in the cosmology, e.g., unified dark energy models.

As an example, we have shown how the generalised Chaplygin gas cosmology can be re-derived as a solution to an interacting matter+vacuum model. In particular we have shown that the interaction can be defined by a single dimensionless constant and the late-time constant vacuum energy (which appears as a dimensional parameter in the original Chaplygin gas model) can instead be derived as a constant of integration in the matter+vacuum model. We leave it to future work to re-assess the cosmological constant problem in this particular model, and to study the constraints placed on the decomposed model from the full range of cosmological observations, including inhomogeneous perturbations and the growth of structure.

\acknowledgments We thank Marco Bruni, Rob Crittenden, Georgia Kittou, Kazuya Koyama and Roy Maartens for helpful comments. DW is supported by STFC grant ST/H002774/1. JD-S is supported by CONACYT grant 210405. JD-S and YW thank the ICG, University of Portsmouth for their hospitality.



\begin{thebibliography}{99}

\bibitem{Copeland:2006wr}
  E.~J.~Copeland, M.~Sami and S.~Tsujikawa,
  Int.\ J.\ Mod.\ Phys.\ D {\bf 15}, 1753 (2006)
  [hep-th/0603057].

\bibitem{Weinberg:1988cp}
  S.~Weinberg,
  Rev.\ Mod.\ Phys.\  {\bf 61}, 1 (1989).

\bibitem{Bertolami:1986bg}
  O.~Bertolami,
  Nuovo Cim.\ B {\bf 93}, 36 (1986).

\bibitem{Freese:1986dd}
  K.~Freese, F.~C.~Adams, J.~A.~Frieman and E.~Mottola,
  Nucl.\ Phys.\ B {\bf 287}, 797 (1987).

\bibitem{Chen:1990jw}
  W.~Chen and Y.~S.~Wu,
  Phys.\ Rev.\ D {\bf 41}, 695 (1990)
  [Erratum-ibid.\ D {\bf 45}, 4728 (1992)].

\bibitem{Carvalho:1991ut}
  J.~C.~Carvalho, J.~A.~S.~Lima and I.~Waga,
  Phys.\ Rev.\ D {\bf 46}, 2404 (1992).

\bibitem{Berman:1991zz}
  M.~S.~Berman,
  Phys.\ Rev.\ D {\bf 43}, 1075 (1991).

\bibitem{Pavon:1991uc}
  D.~Pavon,
  Phys.\ Rev.\ D {\bf 43}, 375 (1991).

\bibitem{AlRawaf:1995rs}
  A.~S.~Al-Rawaf and M.~O.~Taha,
  Phys.\ Lett.\ B {\bf 366}, 69 (1996).

\bibitem{Shapiro:2000dz}
  I.~L.~Shapiro and J.~Sola,
  JHEP {\bf 0202}, 006 (2002)
  [hep-th/0012227].


\bibitem{Sola:2011qr}
  J.~Sola,
  J.\ Phys.\ Conf.\ Ser.\  {\bf 283}, 012033 (2011)
  [arXiv:1102.1815 [astro-ph.CO]].

\bibitem{Bunch:1978yq}
  T.~S.~Bunch and P.~C.~W.~Davies,
  Proc.\ Roy.\ Soc.\ Lond.\ A {\bf 360}, 117 (1978).

\bibitem{Schutzhold:2002pr}
  R.~Schutzhold,
  Phys.\ Rev.\ Lett.\  {\bf 89}, 081302 (2002).

\bibitem{Fabris:2006gt}
  J.~C.~Fabris, I.~L.~Shapiro and J.~Sola,
  JCAP {\bf 0702}, 016 (2007)
  [gr-qc/0609017].

\bibitem{VelasquezToribio:2009qp}
  A.~M.~Velasquez-Toribio,
  Int.\ J.\ Mod.\ Phys.\ D {\bf 21}, 1250026 (2012)
  [arXiv:0907.3518 [astro-ph.CO]].

\bibitem{Zimdahl:2011ae}
  W.~Zimdahl, H.~A.~Borges, S.~Carneiro, J.~C.~Fabris and W.~S.~Hipolito-Ricaldi,
  JCAP {\bf 1104}, 028 (2011)
  [arXiv:1009.0672 [astro-ph.CO]].

\bibitem{Xu:2011qv}
  L.~Xu, Y.~Wang, M.~Tong and H.~Noh,
  Phys.\ Rev.\ D {\bf 84}, 123004 (2011)
  [arXiv:1112.5216 [astro-ph.CO]].

\bibitem{Alcaniz:2012mh}
  J.~S.~Alcaniz, H.~A.~Borges, S.~Carneiro, J.~C.~Fabris, C.~Pigozzo and W.~Zimdahl,
  arXiv:1201.5919 [astro-ph.CO].

\bibitem{Ratra:1987rm}
  B.~Ratra and P.~J.~E.~Peebles,
  Phys.\ Rev.\ D {\bf 37}, 3406 (1988).

\bibitem{Caldwell:1997ii}
  R.~R.~Caldwell, R.~Dave and P.~J.~Steinhardt,
  Phys.\ Rev.\ Lett.\  {\bf 80}, 1582 (1998)
  [astro-ph/9708069].

\bibitem{ArmendarizPicon:2000ah}
  C.~Armendariz-Picon, V.~F.~Mukhanov and P.~J.~Steinhardt,
  Phys.\ Rev.\ D {\bf 63}, 103510 (2001)
  [astro-ph/0006373].

\bibitem{Malik:2004tf}
  K.~A.~Malik and D.~Wands,
  JCAP {\bf 0502}, 007 (2005)
  [astro-ph/0411703].

\bibitem{Wetterich:1994bg}
  C.~Wetterich,
  Astron.\ Astrophys.\  {\bf 301}, 321 (1995)
  [hep-th/9408025].

\bibitem{Amendola:1999er}
  L.~Amendola,
  Phys.\ Rev.\ D {\bf 62}, 043511 (2000)
  [astro-ph/9908023].

\bibitem{Holden:1999hm}
  D.~J.~Holden and D.~Wands,
  Phys.\ Rev.\ D {\bf 61}, 043506 (2000)
  [gr-qc/9908026].

\bibitem{Billyard:2000bh}
  A.~P.~Billyard and A.~A.~Coley,
  Phys.\ Rev.\ D {\bf 61}, 083503 (2000)
  [astro-ph/9908224].

\bibitem{Farrar:2003uw}
  G.~R.~Farrar and P.~J.~E.~Peebles,
  Astrophys.\ J.\  {\bf 604}, 1 (2004)
  [astro-ph/0307316].

\bibitem{Boehmer:2008av}
  C.~G.~Boehmer, G.~Caldera-Cabral, R.~Lazkoz and R.~Maartens,
  Phys.\ Rev.\ D {\bf 78}, 023505 (2008)
  [arXiv:0801.1565 [gr-qc]].

\bibitem{Kodama:1985bj}
  H.~Kodama and M.~Sasaki,
  Prog.\ Theor.\ Phys.\ Suppl.\  {\bf 78}, 1 (1984).

\bibitem{Mukhanov:1990me}
  V.~F.~Mukhanov, H.~A.~Feldman and R.~H.~Brandenberger,
  Phys.\ Rept.\  {\bf 215}, 203 (1992).

\bibitem{Malik:2008im}
  K.~A.~Malik and D.~Wands,
  Phys.\ Rept.\  {\bf 475}, 1 (2009)
  [arXiv:0809.4944 [astro-ph]].

\bibitem{ArmendarizPicon:1999rj}
  C.~Armendariz-Picon, T.~Damour and V.~F.~Mukhanov,
  Phys.\ Lett.\ B {\bf 458}, 209 (1999)
  [hep-th/9904075].

\bibitem{DeSantiago:2012inprep}
  J.~De-Santiago, J.~L.~Cervantes-Cota and D.~Wands,
  {\em in preparation}.

\bibitem{Scherrer:2004au}
  R.~J.~Scherrer,
  Phys.\ Rev.\ Lett.\  {\bf 93}, 011301 (2004)
  [astro-ph/0402316].

\bibitem{Garriga:1999vw}
  J.~Garriga and V.~F.~Mukhanov,
  Phys.\ Lett.\ B {\bf 458}, 219 (1999)
  [hep-th/9904176].

\bibitem{Kamenshchik:2001cp}
  A.~Y.~.Kamenshchik, U.~Moschella and V.~Pasquier,
  Phys.\ Lett.\ B {\bf 511}, 265 (2001)
  [gr-qc/0103004].

\bibitem{Linder:2008ya}
  E.~V.~Linder and R.~J.~Scherrer,
  Phys.\ Rev.\ D {\bf 80}, 023008 (2009)
  [arXiv:0811.2797 [astro-ph]].

\bibitem{Kunz:2007rk}
  M.~Kunz,
  Phys.\ Rev.\ D {\bf 80}, 123001 (2009)
  [astro-ph/0702615].

\bibitem{Aviles:2011ak}
  A.~Aviles and J.~L.~Cervantes-Cota,
  Phys.\ Rev.\ D {\bf 84}, 083515 (2011)
  [Erratum-ibid.\ D {\bf 84}, 089905 (2011)]
  [arXiv:1108.2457 [astro-ph.CO]].

\bibitem{Bento:2002ps}
  M.~C.~Bento, O.~Bertolami and A.~A.~Sen,
  Phys.\ Rev.\ D {\bf 66}, 043507 (2002)
  [gr-qc/0202064].

\bibitem{Sandvik:2002jz}
  H.~Sandvik, M.~Tegmark, M.~Zaldarriaga and I.~Waga,
  Phys.\ Rev.\ D {\bf 69}, 123524 (2004)
  [astro-ph/0212114].

\bibitem{Park:2009np}
  C.~-G.~Park, J.~-c.~Hwang, J.~Park and H.~Noh,
  Phys.\ Rev.\ D {\bf 81}, 063532 (2010)
  [arXiv:0910.4202 [astro-ph.CO]].

\bibitem{Bento:2004uh}
  M.~C.~Bento, O.~Bertolami and A.~A.~Sen,
  Phys.\ Rev.\ D {\bf 70}, 083519 (2004)
  [astro-ph/0407239].

\bibitem{Barrow:2006hia}
  J.~D.~Barrow and T.~Clifton,
  Phys.\ Rev.\ D {\bf 73}, 103520 (2006)
  [gr-qc/0604063].

\end{thebibliography}
\end{document}